\documentclass[aps,prr,reprint,superscriptaddress]{revtex4-1}
\usepackage{bm,siunitx}
\usepackage{amsmath}
\DeclareSIUnit\gauss{G}
\usepackage{graphicx}
\draft 
\usepackage{color}

\begin{document}


\title{Loading a quantum gas from a hybrid dimple trap to a shell trap} 

\author{David Rey}
\author{Simon Thomas}
\author{Rishabh Sharma}
\author{Thomas Badr}
\author{Laurent Longchambon}
\author{Romain Dubessy}
\author{Hélène Perrin}
\email[]{helene.perrin@univ-paris13.fr}
\affiliation{Laboratoire de Physique des Lasers, Université Sorbonne Paris Nord, CNRS UMR 7538, F-93430, Villetaneuse, France}

\date{\today}

\begin{abstract}
Starting from a degenerate Bose gas in a hybrid trap combining a magnetic quadrupole trap and an attractive optical trap resulting from a focused laser beam, we demonstrate the efficient loading of this quantum gas into a shell-shaped trap. The shell trap is purely magnetic and relies on adiabatic potentials for atoms in an inhomogeneous magnetic field dressed by a radiofrequency (rf) field. We show that direct rf evaporation in the hybrid trap enables an efficient and simple preparation of the cold sample, well adapted to the subsequent loading procedure. The transfer into the shell trap is adiabatic and limits the final excitation of the center-of-mass motion to below \SI{2}{\micro\metre}.
\end{abstract}


\maketitle 

\section{Introduction}
\label{sec:intro}
Cold atoms, or quantum gases when cooled to below the quantum degeneracy regime, represent a powerful tool for applied physics: they are instrumental in precision measurements \cite{Cronin2009} or clocks \cite{Ludlow2015}; they have opened the field of quantum simulation with quantum gases \cite{Bloch2012,AmicoAtomtronics} or collections of singly trapped atoms in Rydberg states \cite{Browaeys2020}. An important feature of cold atoms is the control over their trapping geometry. This includes harmonic traps obtained by magnetic or optical potentials \cite{Ketterle1999,Grimm2000}, box traps \cite{Gaunt2013}, lattices \cite{Bloch2005}, double-wells \cite{Schumm2005b}, rings \cite{Ryu2007}, bubble- or shell-traps \cite{Merloti2013a} or even arbitrary potentials using tailored light \cite{Gauthier2021}.

In particular, traps where atoms are confined to a thin shell, obtained using the adiabatic potentials resulting from radiofrequency (rf) dressing of atoms in an inhomogeneous magnetic field \cite{Garraway2016,Perrin2017}, have developed widely in the recent years. Taking advantage of the strong anisotropy of the confinement, they have been used to study two-dimensional Bose gases gathered at the bottom of the shell in the presence of gravity \cite{Merloti2013a,DeRossi2016,Barker2020,Sunami2022}. The anharmonic nature of the potential has also been exploited to prepare strongly out of equilibrium superfluid samples flowing beyond the speed of sound \cite{Guo2020}. On the other hand, filling the full shell with a quantum gas gives access to a different topology and has motivated an experiment in the microgravity of the International Space Station \cite{Carollo2022}, and an experiment on the Earth where gravity compensation has enabled the preparation of an annular gas \cite{Guo2022}.

Loading a quantum gas into such a shell trap is not straightforward. Quantum gases are usually prepared from a three-dimensional harmonically trapped sample by evaporative cooling \cite{Ketterle1999} and then loaded into the shell trap. As adiabatic potentials for rf-dressed atoms rely on the use of inhomogeneous magnetic fields, the initial trap is generally a purely magnetic trap \cite{Colombe2004a,Schumm2005b,Gildemeister2010} or a quadrupole trap plugged in its center with a blue-detuned laser beam to prevent losses in the region of low magnetic fields \cite{Davis1995,Naik2005}. While we have used this configuration in our previous experiments with shell traps, the loading process from the plugged quadrupole trap to the shell trap requires a delicate procedure to ramp down the laser beam in order to prevent excitations of the center of mass after switching the light \cite{Merloti2013a}. This is related to the fact that the initial cloud is located on one side of the quadrupole center, whereas the final cloud sits just below, requiring a displacement of the cloud both in vertical and horizontal directions. The excitations that appear when the alignment is not optimum can be problematic, especially for the study of the superfluid dynamics of a quantum gas, which is one of the situations where the adiabatic potentials are otherwise ideally appropriate because of their very small roughness \cite{Guo2020,Pandey2019}.

\begin{figure}
    \centering
    \includegraphics[width=8.6cm,clip=true,trim=0mm 20mm 0mm 0mm]{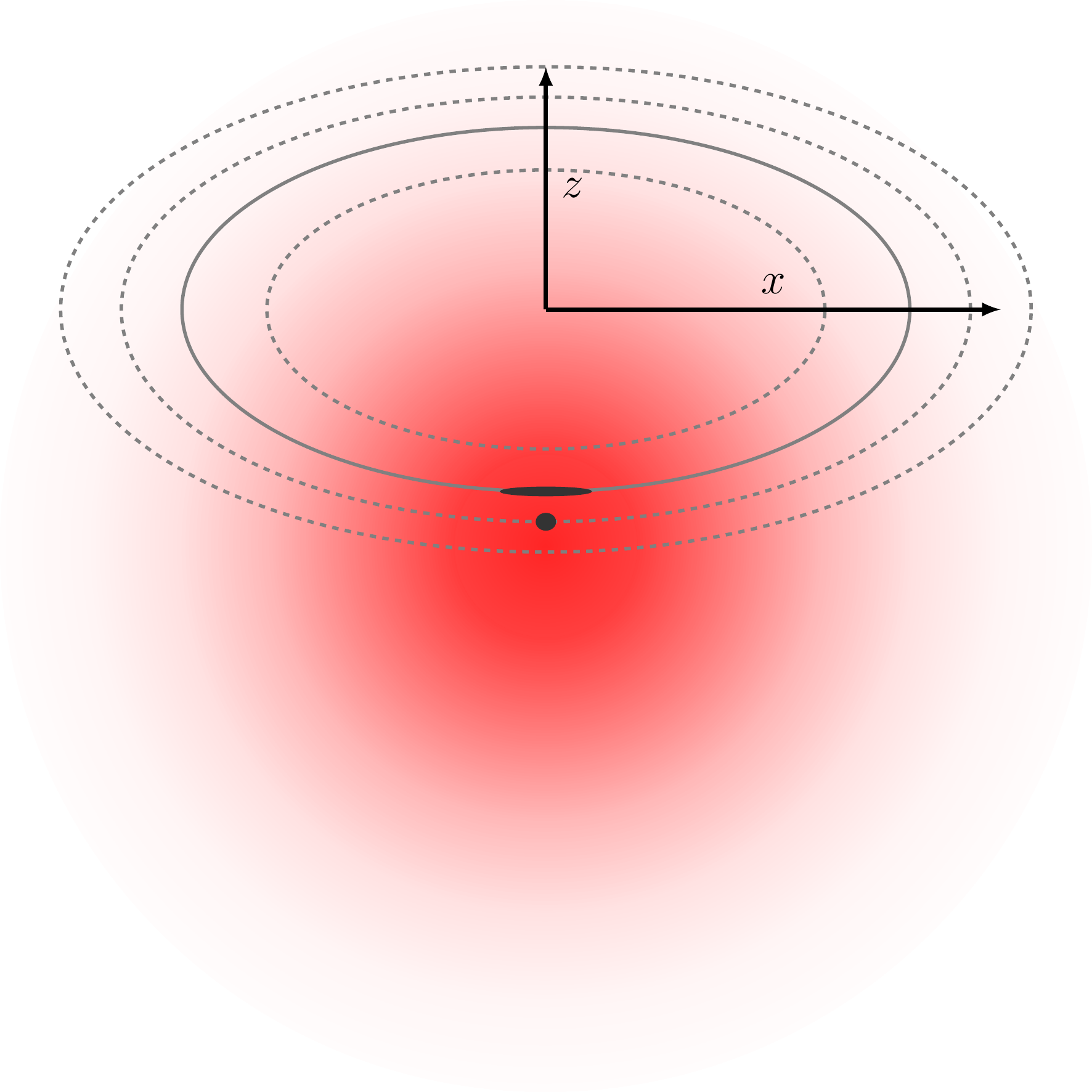}
    \caption{Sketch of the transfer of the quantum gas from the hybrid trap (lower gray dot) to the shell trap (gray ellipse). The $x$ and $z$ axes are centered to the zero of the magnetic quadrupole. The red-detuned laser is represented as well at the isomagnetic surfaces corresponding, from outermost to innermost, to the end of rf evaporation at \SI{400}{\kilo\hertz}, the center of the hybrid trap, the final shell (full black line) at \SI{300}{\kilo\hertz} and the initial rf frequency of the loading ramp at \SI{230}{\kilo\hertz}.}
    \label{fig:trap}
\end{figure}

In this paper, we propose and demonstrate a new approach to prepare a quantum gas in a shell trap. We use a hybrid trap \cite{Lin2009b}, combining a magnetic quadrupole field and a red-detuned `dimple' laser beam, focused below the zero of the quadrupole, approximately at the position where the atoms will be finally confined in the shell trap, see Fig.~\ref{fig:trap}. This strategy minimizes the displacement of the cloud during the transfer from one trap to the other, hence its subsequent excitation. We show that the preparation of the quantum gas is efficient and robust to small misalignment.

The paper is organized as follows. After describing in some detail the production of the BEC in a hybrid trap in Sec.~\ref{sec:hybrid_trap}, we will describe the loading process to the shell trap in Sec.~\ref{sec:loading} and give its performances. We end with a final discussion in Sec.~\ref{sec:discussion}.

\section{Hybrid dimple trap}
\label{sec:hybrid_trap}
\subsection{Adapting the hybrid trap to the final shell trap}
As the hybrid dimple trap is used as an intermediate step for the preparation of a quantum gas before the final shell trap, we first discuss the characteristics of the latter in order to adapt the parameters of the former for an optimal transfer.

The shell-shaped trap is obtained by dressing with an rf field of frequency $\omega$ atoms of total spin $F$ placed in an inhomogeneous magnetic field \cite{Zobay2001,Merloti2013a,Perrin2017}. If the coupling amplitude $\Omega_0$ between the spin and the rf field is strong enough, the atoms follow at each point of space an eigenstate of the Hamiltonian coupling its spin with the static and rf magnetic fields, thus experiencing an adiabatic potential. The eigenstates present an avoided crossing on the surfaces defined by $\omega_0(\mathbf{r})=\omega$, where the rf field is resonant with the local Zeeman splitting $\omega_0(\mathbf{r})$ between magnetic states. For atoms in the upper dressed state, this avoided crossing represents a minimum of their adiabatic potential.

In our experiment, the underlying inhomogeneous magnetic field is a quadrupole field of symmetry axis $z$, leading to a Zeeman splitting of the form
\[
\omega_0(\mathbf{r}) = \alpha\sqrt{x^2+y^2+4z^2} = \alpha\ell(\mathbf{r})
\]
where $\alpha$ is the magnetic gradient in units of frequency and $\ell(\bm{r})=\sqrt{x^2+y^2+4z^2}$. Given the polarization of the rf dressing field, circular with respect to the vertical axis $z$, the adiabatic potential for atoms in the upper substate of $F=1$ is given by \cite{Merloti2013a}
\begin{equation}
V_{\rm adia}(\bm{r})=\hbar\sqrt{(\alpha\ell(\bm{r})-\omega)^2+\Omega(\bm{r})^2},
\label{eq:pot_adia}
\end{equation}
where $\Omega(\bm{r})=\Omega_0/2\times(1-2z/\ell(\bm{r}))$.

With this quadrupole field, the resonant surfaces are ellipsoids, with a main radius $r_0=\omega/\alpha$, see Fig.~\ref{fig:trap}. In the presence of gravity, atoms confined in the shell trap gather at the bottom of this ellipsoid, at position $z_0=-r_0/2$. Therefore, this is the position around which we want to prepare a quantum gas by evaporation in the hybrid trap, in order to minimize the excitations when transferring the atoms from one trap to the other.

In the experiment, we work with a typical rf frequency of \SI{300}{\kilo\hertz} and a gradient $\alpha=2\pi\times\SI{4.14}{\kilo\hertz\per\micro\metre}$ such that $r_0=\SI{72.5}{\micro\metre}$ and $-r_0/2=\SI{-36.2}{\micro\metre}$. As the minimum of the hybrid trap is shifted by the magnetic gradient toward the center of the quadrupole, the red-detuned laser beam should be focused at a position $z_L$ a few micrometers below $z_0$. These considerations have guided the choice of beam position and waist presented in Sec.~\ref{sec:setup}.

\subsection{Experimental setup}
\label{sec:setup}
The experimental setup consists of three vacuum chambers described in Ref. \onlinecite{Dubessy2012a}. Briefly, after producing a cold sample of rubidium 87 atoms in the magneto-optical trap (MOT) chamber, the atoms are magnetically transported toward the science chamber, a glass cell allowing to optically access the cloud for trapping and imaging. In the science cell, they are confined in a magnetic quadrupole trap. The magnetic gradient is produced by two coils of common vertical axis $z$ and can reach a maximum of $b'= \SI{229}{\gauss\per\centi\metre}$ in the horizontal directions $x$ and $y$. The experimental sequence is driven by analog and digital input/output cards from National Instruments\footnote{One PXIe-6738 analog I/O card and two PXIe-6535 digital I/O cards from National Instruments.} installed in a PXI rack, controlled with a temporal resolution at the microsecond level by the Labscript Suite software \cite{Starkey2013,StarkeyThese}. The software contains an optimization package (mloop) \cite{Wigley2016}, relying on Gaussian process regression, that we used at an early stage of the experiment to explore the sensitivity of the final condensed atom number to various parameters of the experimental sequence, including e.g. the duration of the molasses or the ramp of the magnetic field to its maximum value \cite{ReyThese}.

In order to prevent Majorana losses caused by the vanishing magnetic field in the center of the quadrupole trap, several strategies can be implemented. They include the addition of a rotating homogeneous field in the case of a TOP trap \cite{Petrich1995}, of a repulsive optical potential focused in the quadrupole center as for plugged traps \cite{Davis1995,Naik2005,Dubessy2012a} or of a red-detuned optical potential attracting the atoms away from the low magnetic field region \cite{Lin2009b}. Here, we have followed the latter strategy.

We now describe the experimental configuration of our hybrid trap, taking the center of the magnetic quadrupole as the origin of positions $(x,y,z)=(0,0,0)$. A red-detuned laser beam with a wavelength of \SI{1064}{\nano\metre} and maximum power on the atoms $P_{\rm max}=\SI{2.74}{\watt}$ propagates along the direction $y$ and is focused by a single lens of focal length $f=\SI{200}{mm}$ to a waist of approximately \SI{70}{\micro\metre}, comparable to Ref. \onlinecite{Lin2009b}, at a vertical position $z_L\simeq-\SI{50}{\micro\metre}$ below the center of the quadrupole, see Fig.~\ref{fig:trap}. The Rayleigh length is larger than \SI{12}{\milli\metre} and the beam astigmatism is small such that we can neglect their effect on the trapping potential, which we write as:
\begin{eqnarray}
    V(\bm{r})&=&\hbar\alpha\sqrt{x^2+y^2+4z^2}+Mgz\nonumber\\
    &+&U_0\,\exp\left[{\displaystyle -2\frac{x^2}{w_x^2}-2\frac{(z-z_L)^2}{w_z^2}}\right].
    \label{eqn:Vtrap}
\end{eqnarray}
In this expression, the first term corresponds to the magnetic quadrupole trap, the second term to gravity, and the last one to the optical dipole potential \cite{Grimm2000} where we allow a possible anisotropy of the beam waists. $\alpha$ is the magnetic gradient in units of frequency for atoms in the $F=1$ electronic groundstate, with a maximum value of \cite{deGoer2021} $\alpha_{\rm max}/(2\pi)=\SI{16}{\kilo\hertz\per\micro\metre}$, $M$ is the atomic mass, $g$ is Earth's gravity acceleration, and $w_x$ and $w_z$ the beam radii at $1/{\rm e}^2$ in the direction $x$ and $z$, respectively. The beam waists deduced from a measurement of the oscillation frequencies of a degenerate gas in the hybrid trap, see below, are $w_x=\SI{72.2}{\micro\metre}$ and $w_z = \SI{61.3}{\micro\metre}$. The optical potential depth $U_0$ scales as\footnote{For $^{87}$Rb and a laser at \SI{1064}{\nano\metre}, the light shift in frequency units is $U_0=qP/(w_xw_z)$ where $q=\SI{2.02}{\giga\hertz\per\watt\square\micro\metre}$. Both $D_1$ and $D_2$ lines contribute to this light shift, as well as both corotating and counterrotating terms \cite{Grimm2000}.} $U_0\propto P/(w_x w_z)$ and is equal to $U_{\rm max}=k_B\times \SI{60}{\micro\kelvin}$ at $P=P_{\rm max}$.

\begin{figure}
    \centering
    \includegraphics[width=8.6cm]{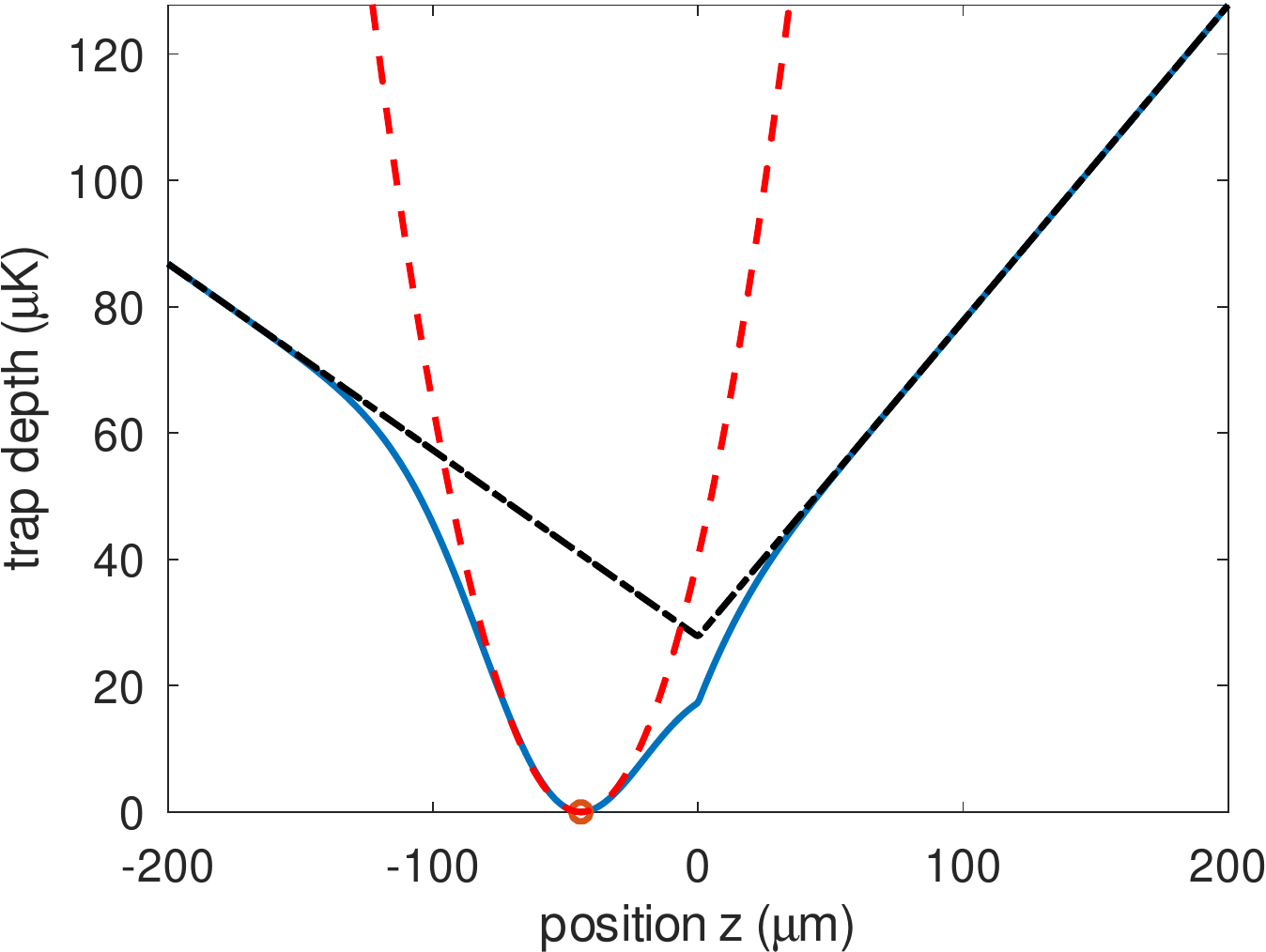}
    \caption{Blue line: Cut in the total potential $V(0,0,z)$ along the $z$ axis, given in units of \SI{}{\micro\kelvin}. Trap parameters: $\alpha=\alpha_{h}=2\pi\times\SI{4.14}{\kilo\hertz\per\micro\metre}$, $P=P_h=\SI{1.92}{\watt}$. Red dashed line: Harmonic approximation near the trap bottom. Black dot-dashed line: Potential resulting from the quadrupole trap and gravity alone.}
    \label{fig:trap_cut_z}
\end{figure}

\subsection{Hybrid trap characterization}
\label{sec:characterization}
We now explain how we characterize the trap parameters. We perform this characterization at the laser power and magnetic gradient used in the last stage of evaporative cooling, see Sec.~\ref{sec:evap_strategy}. The total power $P_h=\SI{1.92\pm0.02}{\watt}$ is known by direct measurement with a powermeter. The magnetic gradient $\alpha_h=2\pi\times\SI{4.14\pm0.06}{\kilo\hertz\per\micro\metre}$ has been measured previously \cite{Guo2022} by recording the position of the cloud in the shell trap when scanning the rf dressing frequency. A cut along $z$ of the corresponding potential is shown in Fig.~\ref{fig:trap_cut_z}.

In order to fully characterize the trap, we need to measure the dimple beam waists $w_x$ and $w_z$ and the position $(x_L=0,y_L=0,z_L)$ of the focus with respect to the zero of the magnetic field $z_L$. $y_L=0$ is ensured by the large Rayleigh length, and we assume a perfect alignment of the beam below the center of the quadrupole $x_L=0$ (see below the alignment procedure). The remaining trap parameters are extracted from three independent measurements: the oscillation frequencies along the axes $\omega_x$ and $\omega_z$ and the position of the atoms in the magnetic field, thanks to rf-spectroscopy.

From Eq.~\eqref{eqn:Vtrap}, we determine the position of trap minimum $(0,0,z_{\rm eq})$ that we use as an intermediate value,
\begin{equation}
z_{\rm eq}\simeq z_L+w_z^2\frac{\hbar\alpha}{2|U_0|}(1-\epsilon),
\label{eq:z_eq}    
\end{equation}
where $\epsilon=Mg/(2\hbar\alpha)=\SI{0.26\pm0.01}{}$ is the gravitational sag parameter \cite{Merloti2013a}. This expression is valid for a shift $z_{\rm eq}- z_L$ small as compared to the beam waist $w_z$. We also extract the oscillation frequencies
\begin{eqnarray}
M\omega_x^2&=&\frac{\hbar\alpha}{2|z_{\rm eq}|}+\frac{4|U_0|}{w_x^2}\,\exp\left[{\displaystyle-2\frac{(z_{\rm eq}-z_L)^2}{w_z^2}}\right],\label{eq:omega_x}\\
M\omega_y^2&=&\frac{\hbar\alpha}{2|z_{\rm eq}|},\label{eq:omega_y}\\
M\omega_z^2&=&\frac{4|U_0|}{w_z^2}\left(1 - 4\frac{(z_{\rm eq}-z_L)^2}{w_z^2}\right)\,{\rm e}^{-2\frac{(z_{\rm eq}-z_L)^2}{w_z^2}}.\label{eq:omega_z}
\end{eqnarray}

Crucially, these formulas assume that the dimple beam is well centered onto the vertical axis, i.e., $x_L=0$. To ensure this we follow an alignment procedure in two steps. The position of the dimple beam in the $(x,z)$ plane is controlled by a piezo-actuated mirror with a sensitivity of about $\SI{1}{\micro\metre\per\volt}$ in the object plane. As the dimple beam is co-linear with our absorption imaging beam we first use a weak leak through a dichroic filter to directly image its \textit{in situ} position and perform a first rough alignment, limited by the optical resolution and aberrations. In a second fine alignment step, we use directly the atoms: we shine an rf-field at $\omega_{\rm rf}/(2\pi)=\SI{400}{kHz}$, strong enough to dress the atoms, and we record the losses induced by a full spin flip to the state $F=1,m_F=1$, which is anti-trapped along the $y$ axis. This occurs when the Zeeman splitting between the sub-levels of the $F=1$ manifold at the position of the atoms is resonant with the applied rf field. By moving around the dimple beam, we are able to locate precisely the resonant ellipsoid $\hbar\alpha\sqrt{x^2+4z^2}=\hbar\omega_{\rm rf}$ and position the beam center with respect to the magnetic field.

With the additional parameter $z_{\rm eq}$, we have four parameters to determine: $z_{\rm eq}$, $z_L$, $w_x$ and $w_z$. The determination of $z_{\rm eq}$ could be done from Eq.~\eqref{eq:omega_y} and the measurement of the oscillation frequency in the $y$ direction $\omega_y$. However, it is more accurate to deduce $z_{\rm eq}$ from an rf spectroscopy measurement of the value of the linear Zeeman splitting\footnote{Here we take into account only the linear Zeeman splitting as the local magnetic field is weak ($\sim\SI{0.5}{G}$) and differential light shifts from the dimple beam are completely negligible (of the order of a few Hz).} in the quadrupole at equilibrium,
\begin{equation}
\omega_0=2\alpha|z_{\rm eq}|.
\label{eq:Larmor}
\end{equation}
In order to get all four parameters, we first determine $z_{\rm eq}=-\omega_0/(2\alpha)$ from the measurement of $\omega_0$. Then, noticing that the term $z_{\rm eq}-z_L$ that appears in Eqs. \eqref{eq:omega_x} and \eqref{eq:omega_z} does not depend on $z_L$,  we determine $w_x$ and $w_z$ numerically from Eqs. \eqref{eq:omega_x} and \eqref{eq:omega_z} and the measurement of $\omega_x$ and $\omega_z$. Finally, we deduce $z_L$ from Eq. \eqref{eq:z_eq}.

The value of $\omega_0$ is obtained from an rf-spectroscopy measurement. We shine a weak rf probe field onto the atoms held inside the hybrid trap during $\SI{500}{ms}$ and scan the rf-frequency. In contrast to the previous case with a strong field, the weak rf field only couples the trapped $m_F=-1$ state to the trapped $m_F=0$ state. As a result, when hitting the resonance we observe no atom losses but heating. This can be understood as an anti-evaporation mechanism: at resonance, the rf-field removes the coldest atoms from $m_F=-1$ and the remaining atoms thermalize to a higher temperature. The atoms transferred to $m_F=0$ remain trapped but can explore a much larger volume in the $y$ direction where the weak confinement is only due to the focusing of the dimple beam\footnote{This residual trapping frequency is estimated to be of the order of $\omega_y/(2\pi)\simeq\SI{1}{Hz}$}.

\begin{figure}
    \centering
    \includegraphics[width=8.6cm]{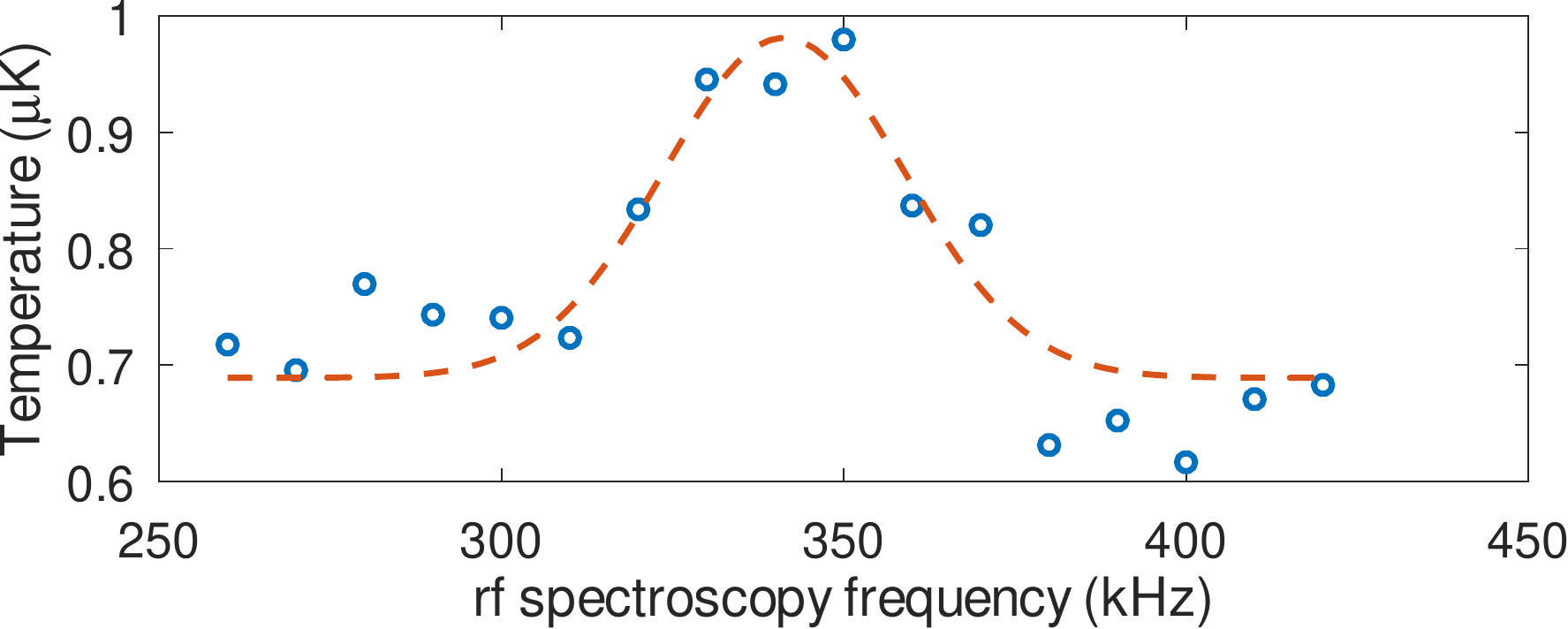}
    \caption{Open circles: Temperature of the atoms in the hybrid trap as a function of the frequency of the weak rf probe. The dashed red line is a fit with a Gaussian model of the resonant behavior. See text for details.}
    \label{fig:spectro_freqy}
\end{figure}

Figure~\ref{fig:spectro_freqy}a) shows such a measurement where a resonance in the temperature is clearly observed while the atom number remains constant at $N=\SI{2.1\pm0.2e5}{}$. From a Gaussian fit of the resonance, we obtain $\omega_0/(2\pi)=\SI{341.3\pm2.4}{kHz}$. This gives us the equilibrium position $z_{\rm eq}=-\SI{41.2\pm0.9}{\micro\metre}$. From $z_{\rm eq}$, we can deduce the oscillation frequency $\omega_y=\alpha\sqrt{\hbar/(M\omega_0)}=2\pi\times\SI{76.4\pm1.4}{\hertz}$, in agreement with direct measurements.

In order to determine the remaining parameters $(w_x,w_z,z_L)$, we measure oscillation frequencies $\omega_x$ and $\omega_z$ and find numerically the solution of the coupled non-linear equations \eqref{eq:z_eq}, \eqref{eq:omega_x} and \eqref{eq:omega_z}. To excite the dipole motion along the $x$ axis, we slowly ramp a small homogeneous bias magnetic field oriented along $x$ in \SI{50}{ms} and then abruptly (in \SI{50}{\micro\second}) switch it off, resulting in center-of-mass oscillations in the horizontal direction. We proceed in the same manner to excite the motion along the $z$ direction. In both cases, we measure the position of the center of mass after a \SI{23}{\milli\second} time-of-flight expansion.

\begin{figure}
    \centering
    \includegraphics[width=8.6cm]{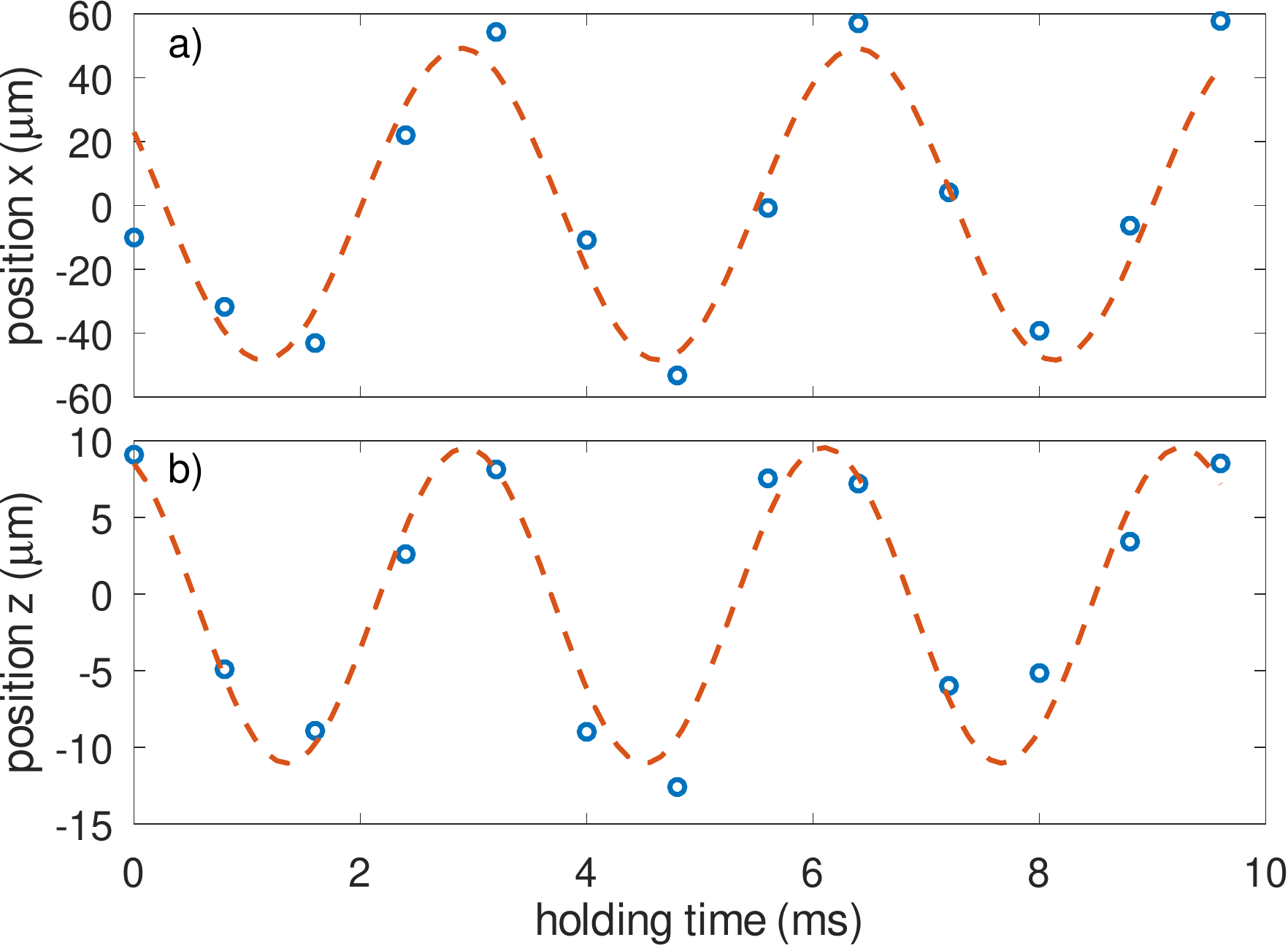}
    \caption{
    Open circles: Dipole oscillations of the center of mass in the hybrid trap after excitation, as measured after a \SI{23}{\milli\second} time-of-flight expansion. a) Along the $x$ axis. b) Along the $z$ axis. For both panels the dashed red curve is a sinusoidal fit to the data. See text for details.
    }
    \label{fig:oscillation_xz}
\end{figure}

Figure~\ref{fig:oscillation_xz}a) and b) display the measurement of the dipole oscillations along the $x$ and $z$ axis, respectively. From a sinusoidal fit of the data, we extract the two oscillation frequencies $\omega_x/(2\pi)=\SI{286.1\pm6.9}{Hz}$ and $\omega_z/(2\pi)=\SI{316.5\pm5.1}{Hz}$ and with a least square fitting procedure we deduce the parameters $w_x=\SI{72.2\pm1.7}{\micro\metre}$ and $w_z=\SI{61.3\pm1.1}{\micro\metre}$. We finally get $z_L=\SI{-48.0\pm1.1}{\micro\metre}$ from Eq. \eqref{eq:z_eq}.
To estimate the uncertainty, we repeat the least square fitting procedure by varying each of the input parameters (oscillation frequencies and rf spectroscopy resonant frequency), one at a time, on a range corresponding to their experimental uncertainty. We find that the result of the fit always depends linearly on each input parameter, with different slopes for $z_L$, $w_x$ and $w_z$. We then compute the estimated total uncertainty by propagating the experimental errors with the appropriate slopes. From this procedure we also obtain that the uncertainty on $z_L$ is dominated by the uncertainty on $\alpha$ and $\omega_z$, while $w_z$ is mainly affected by the uncertainty on $\omega_z$ and both $\omega_x$ and $\omega_z$ contribute equally to the uncertainty on $w_x$.

\section{Evaporation in the hybrid trap and transfer to the shell trap}
In this section, we report on the evaporative cooling strategy that we implement in view of the subsequent transfer to the shell trap, which we discuss in Sec.~\ref{sec:loading}.
\subsection{Evaporative cooling strategy}
\label{sec:evap_strategy}
The preparation of a quantum gas in the hybrid trap is done in three phases. First, radiofrequency (rf) evaporation is performed in the quadrupole trap at maximum gradient $\alpha_{\rm max}$, with the laser beam at maximum power. During this first phase, the temperature is much larger than the depth of the optical potential and the influence of the laser beam is limited. We follow the same evaporation sequence as in the case of a quadrupole trap plugged with a blue-detuned laser beam \cite{Dubessy2012a}, ramping the rf frequency from \SI{65}{\mega\hertz} down to \SI{4}{\mega\hertz} in \SI{13.6}{\second}.

At the end of this first step, Majorana spin flips start to play a role. The second step consists in lowering the magnetic gradient from $\alpha_{\rm max}$ down to $\alpha_{h}=2\pi\times\SI{4.14}{\kilo\hertz\per\micro\metre}$ in \SI{50}{\milli\second} to reduce the density near the magnetic zero and prevent losses. In the meantime, the laser power is ramped down from $P_{\rm max}$ to $P_h=\SI{1.92}{\watt}$ to limit scattering of spontaneous photons, which decreases proportionally from \SI{0.22}{\per\second} down to \SI{0.15}{\per\second}.
After this opening procedure, the trap frequencies are\footnote{These values differ slightly from the values of Sec.~\ref{sec:characterization} because the laser beam position $z_L=\SI{-51}{\micro\metre}$ as deduced from the rf spectroscopy was slightly different.} $(285,74,317)$\,Hz, and we obtain a cold sample of about $10^7$ atoms at $T\simeq\SI{20}{\micro\kelvin}$ and a peak phase space density (PSD) below $10^{-4}$. A cut of the potential $V(0,0,z)$ along the $z$ axis with these parameters is represented in Fig.~\ref{fig:trap_cut_z}.

The last step of evaporation leads to condensation in the hybrid trap. In Ref. \onlinecite{Lin2009b}, this is done by lowering the magnetic gradient further to perform direct optical evaporation in the dimple trap. This can also be done in our experiment, leading to a condensate after typically \SI{6}{\second} of optical evaporation down to 5\%--10\% of the total laser power. However, the magnetic gradient, in this case, is very low, which is not appropriate for the final loading of the shell trap: the gradient must be brought back to its initial value to start the loading procedure, which takes another \SI{2.7}{\second} and results in additional complexity and possible heating of the cloud. We have implemented instead a new strategy, the direct rf evaporation in the hybrid trap, at fixed gradient $\alpha_h$ and beam power $P_h$. This method allows us to prepare a pure Bose--Einstein condensate with $N=\SI{1.29\pm0.07e5}{}$ atoms, suitable to be loaded in the shell trap. Below, we explain how we calibrate the trap parameters and describe the evolution of the thermodynamic properties during this last evaporation phase.

\subsection{Direct rf evaporation performances}
We now report on the study of an alternative approach to the last step of condensation, namely direct rf evaporation in the hybrid dimple plus quadrupole trap using the trap parameters determined experimentally. Immediately after the trap opening mentioned in Sec.~\ref{sec:evap_strategy}, we proceed with rf evaporation and ramp the rf frequency $\omega'$ from \SI{2}{MHz} down to \SI{400}{kHz} in \SI{7}{s}. The rf field is strong enough to allow atoms initially in the trapped $m_F=-1$ state to be transferred adiabatically into the anti-trapped $m_F=1$ state provided they reach the evaporation surface. It therefore effectively truncates the trap at finite energy, allowing for efficient rf evaporation. With our parameters, the trap saddle point during evaporation is on the $y$ axis, along which the trapping potential is purely magnetic, therefore ensuring that the trap depth is simply set by $\hbar(\omega^\prime-2\alpha|z_{\rm eq}|)$, i.e., the excess of magnetic energy with respect to the trap minimum. To monitor the efficiency of the process, we record the total atom number $N$ and the temperature $T$ during the rf ramp.

\begin{figure}
    \centering
    \includegraphics[width=8.6cm]{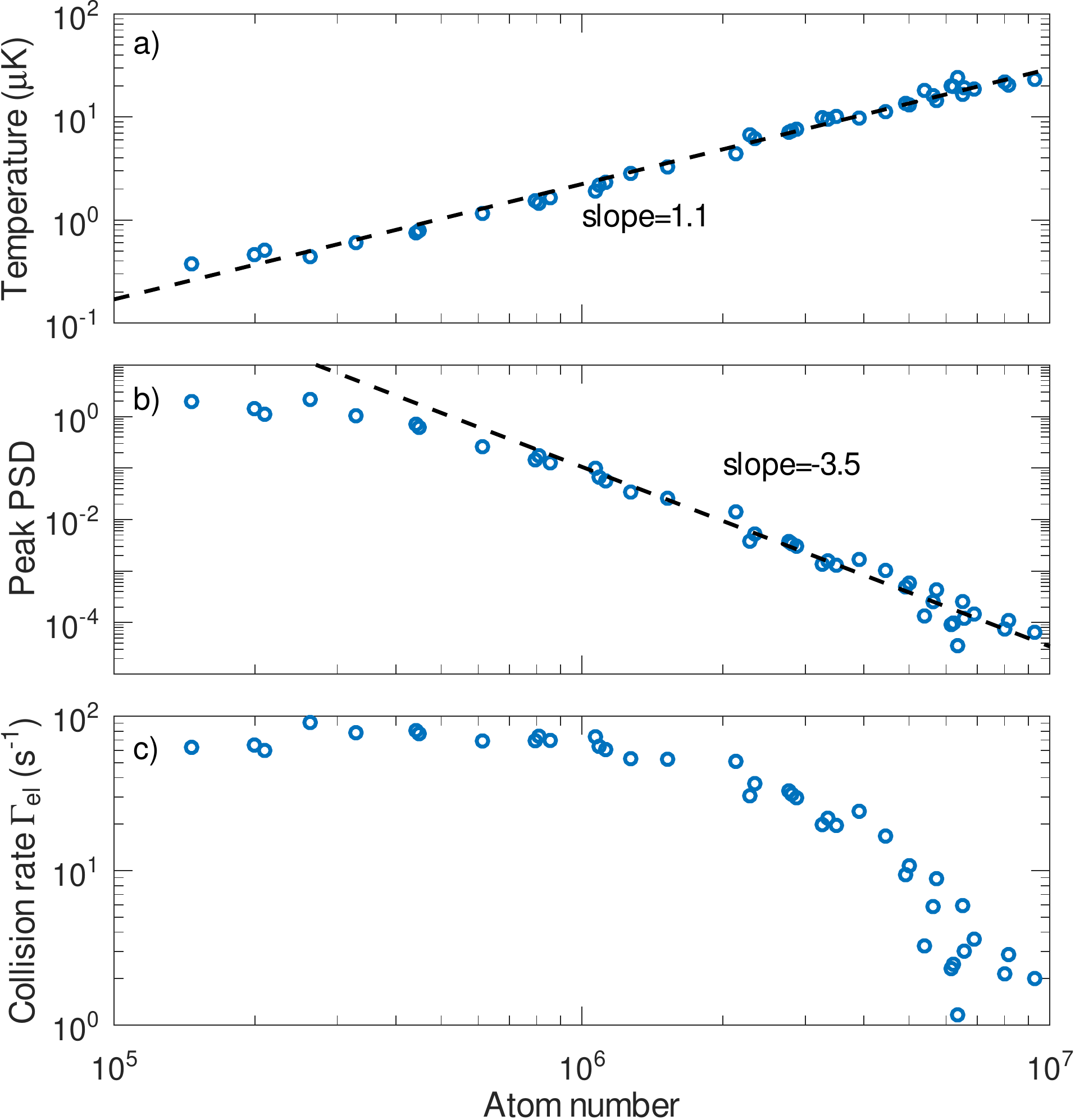}
    \caption{a) Temperature, b) peak phase space density and c) peak collision rate as a function of the atom number during the evaporation ramp (open circles), in log--log scale. For panels a) and b) the dashed black line is a power law fit to the data for atom numbers larger than $10^6$. See text for details.}
    \label{fig:rfscaling}
\end{figure}

Figure~\ref{fig:rfscaling}a) shows that the temperature depends on the atom number with a power law, which is usually observed during evaporative cooling \cite{Luiten1996,Ketterle1996a}, with an exponent here of 1.1. Using our knowledge of the trap potential, we may compute numerically the effective trap volume \cite{Luiten1996}, defined as
\begin{equation}
\mathcal{V}(T)=\int {\rm d}\bm{r}\,\exp{\left(-\frac{V(\bm{r})-V_{\rm eq}}{k_BT}\right)},
\label{eq:effective_volume}
\end{equation}
where $k_B$ is the Boltzmann constant and $V_{\rm eq}=V(0,0,z_{\rm eq})\equiv \textrm{min}[V(\bm{r})]$. Then, using the experimentally measured atom number and temperature, we compute the peak phase space density (PSD) at the center of the trap, $\mathcal{D}=N\Lambda^3/\mathcal{V}(T)$, where $\Lambda=h/\sqrt{2\pi Mk_BT}$ is the thermal de Broglie wavelength.
Figure~\ref{fig:rfscaling}b) displays the evolution of the peak PSD during the evaporation. For $N<4\times10^5$, the PSD reaches the condensation threshold and a condensate appears in the time-of-flight distribution that becomes bimodal. At large atom numbers, a power-law scaling is seen, with a slope of $-3.5$.

The collision rate, shown in Fig.~\ref{fig:rfscaling}c), increases during the first part of evaporation, a signature of the run-away regime, then stays constant when the atom number gets below $N=10^6$. This change in behavior is likely related to the fact that the preferential evaporation along the $y$ axis becomes more pronounced as the total energy of the cloud is reduced, making the evaporation process transit from a three-dimensional to a one-dimensional regime.

\subsection{Loading the shell trap}
\label{sec:loading}
Finally, we demonstrate the loading of an rf-dressed shell-shaped adiabatic potential \cite{Merloti2013a} directly from the dimple trap using a simple and robust procedure.
After switching off the rf field used for evaporation, we turn on an rf dressing field of circular polarization along $z$ produced by three orthogonal antennas \cite{Merloti2013a,Guo2022}, with a coupling amplitude of $\Omega_0/(2\pi)=\SI{50}{\kilo\hertz}$. The initial frequency is $\omega_i/(2\pi)=\SI{230}{\kilo\hertz}$, corresponding to the inner dashed ellipse of Fig.~\ref{fig:trap}, below the resonant frequency at the location of the atoms. We then ramp linearly the rf frequency up to a value of $\omega_f/(2\pi)=\SI{300}{\kilo\hertz}$ in \SI{150}{ms} while simultaneously ramping down the dimple trap power. This protocol transfers the atomic cloud into a pure shell-shaped magnetic trap combining static and rf fields. The atoms sit at the bottom of the ellipsoidal shell, as sketched in Fig.~\ref{fig:trap}. During the transfer the atoms undergo a smooth expansion from the hybrid trap of frequencies $(285,74,317)$\,Hz (lower cloud of Fig.~\ref{fig:trap}) to a pancake-shaped trap $(34,34,378)$\,Hz (upper cloud of Fig.~\ref{fig:trap}).

\begin{figure}
    \centering
    \includegraphics[width=8.6cm]{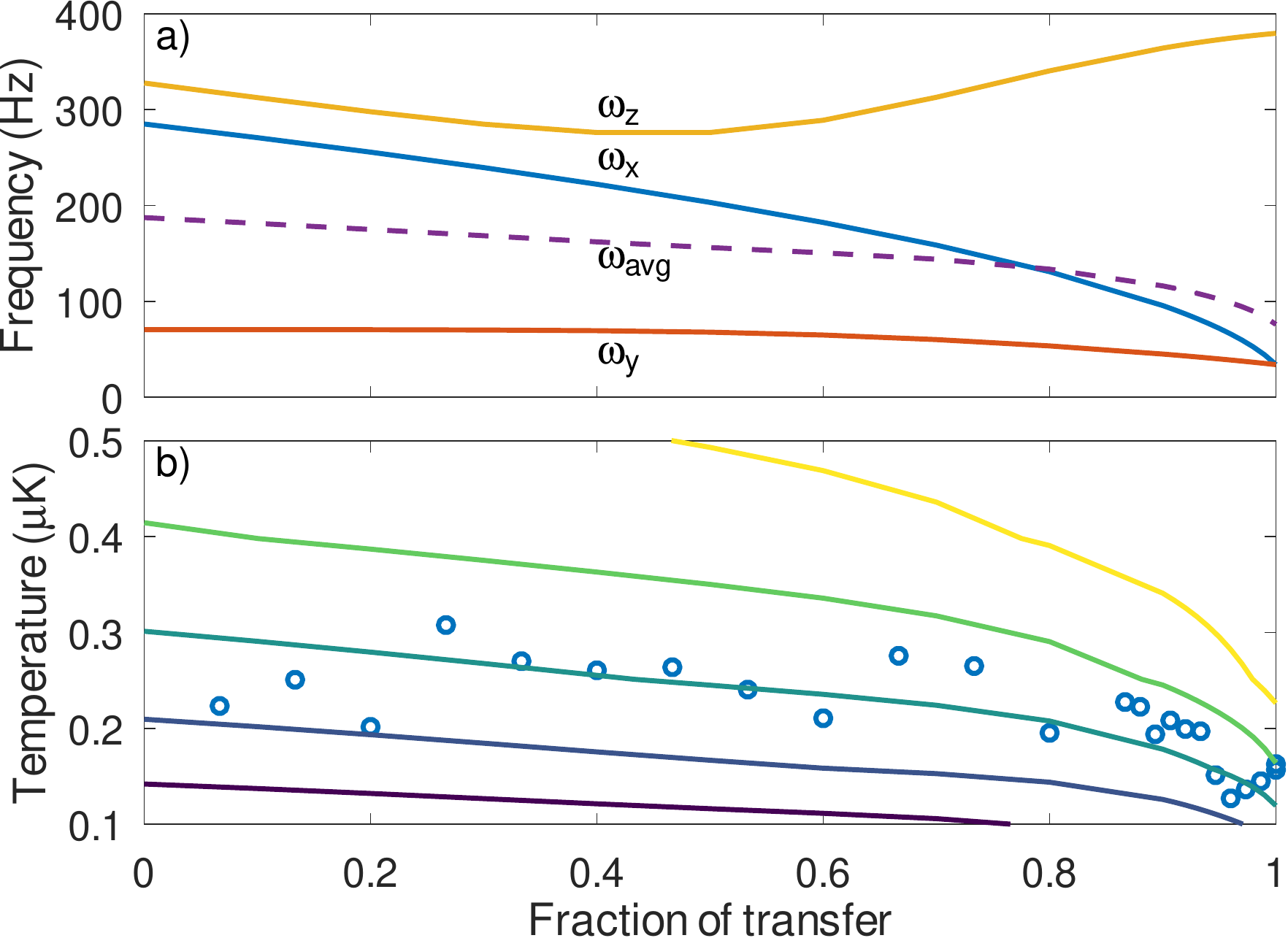}
    \caption{a) Oscillation frequencies $(\omega_x,\omega_y,\omega_z)/(2\pi)$ (blue, red, and yellow solid curves), mean oscillation frequency $\omega_{\rm avg}=(\omega_x\omega_y\omega_z)^{1/3}$ (purple dashed curve), and b) measured temperature (open circles), during the trap transfer. In b), the solid curves correspond to constant entropy trajectories during the transfer, see text for details.}
    \label{fig:dressing}
\end{figure}

Figure~\ref{fig:dressing}a) shows the smooth evolution of the trap frequencies during the transfer. As the average frequency $\omega_{\rm avg}=(\omega_x\omega_y\omega_z)^{1/3}$ decreases, we expect an adiabatic transfer to induce a decrease of the atomic cloud temperature. Figure~\ref{fig:dressing}b) displays the evolution of the measured temperature during the transfer, compared to adiabatic trajectories corresponding to different initial temperatures, see below. We find that the reduction of temperature during the transfer is compatible with an adiabatic trajectory.
We do not measure significant atom losses during the transfer and manage to load $N=\SI{1.8\pm0.2e5}{}$ atoms in the shell trap, at a temperature of \SI{96\pm10}{\nano\kelvin} with a condensate fraction of \SI{46\pm6}{\percent}. To further reduce the final temperature and control the condensate fraction we can add an extra rf knife during the transfer.

To find the isentropic trajectories of Fig.~\ref{fig:dressing}b), we first compute the free energy $F=-Nk_BT\ln{(\mathcal{V}(T)/\Lambda^3)}$ and then the entropy $S=-\partial F/\partial T$. A straightforward computation gives
\begin{equation}
\frac{S}{Nk_B}=\ln{\frac{\mathcal{V}(T)}{\Lambda^3}}+\frac{3}{2}+\frac{\bar{V}}{k_BT},
\label{eqn:entropy}
\end{equation}
where $\bar{V}=\mathcal{V}(T)^{-1}\int {\rm d}\bm{r}\,(V(\bm{r})-V_{\rm eq})\exp{\left(-\frac{V(\bm{r})-V_{\rm eq}}{k_BT}\right)}$ is the average potential energy per atom. To describe the trap potential during dressing, it is sufficient to replace in Eq.~\eqref{eqn:Vtrap} the magnetic quadrupole potential by the adiabatic potential of Eq.~\eqref{eq:pot_adia}, resulting in
\[
V(\bm{r})=\hbar\sqrt{(\alpha\ell-\omega)^2+\Omega(\bm{r})^2}+U_0\,{\rm e}^{-2\frac{x^2}{w_x^2}-2\frac{(z-z_L)^2}{w_z^2}}+Mgz.
\]
Using Eq.~\eqref{eqn:entropy}, it is then possible to find the constant entropy curves for the temperature when the trap geometry is changed from the hybrid trap to the shell trap geometry.

\section{Discussion}
\label{sec:discussion}

As presented in Sec.~\ref{sec:evap_strategy}, we have implemented a direct rf evaporation inside the hybrid dimple plus quadrupole trap, instead of the more usual optical evaporation. This choice is dictated by the subsequent transfer to an rf dressed adiabatic potential, for which we need to maintain a relatively high magnetic gradient, which prevents optical evaporation. Nevertheless, we have found that it is possible to first create a condensate by optical evaporation at a low magnetic gradient and then re-compress the trap by increasing the dimple power and the magnetic gradient, before dressing the atoms with rf photons. However, this strategy offers no improvement and results in a more complex sequence, justifying the interest of direct rf-evaporation.

We have carefully studied this direct evaporation process and performed several measurements to characterize the potential. We have found that the day to day operation is quite robust and does not require a daily accurate characterization of the trap parameters. Indeed, the efficiency of the rf evaporation is mainly set by the trap volume which is not very sensitive to the precise alignment of the laser beam. In particular, even if the beam is slightly off-centered in the horizontal ($x$) direction, the rf evaporation works well. The relevant element is rather the competition between atom loss or heating rates of the cloud and the elastic collision rate. The latter depends mainly on the trap volume, whose order of magnitude is set by the dimple beam power, waist, and magnetic gradient, while the former are controlled by the distance to the magnetic field zero, where Majorana losses occur~ \cite{Naik2005,Dubessy2012a}. Therefore, it is sufficient to maximize the atom number and minimize the temperature at the end of the evaporation ramp by moving around the dimple beam position, at fixed magnetic gradient, to re-align the laser beam and obtain a condensate.

\begin{figure}
    \centering
    \includegraphics[width=8.6cm]{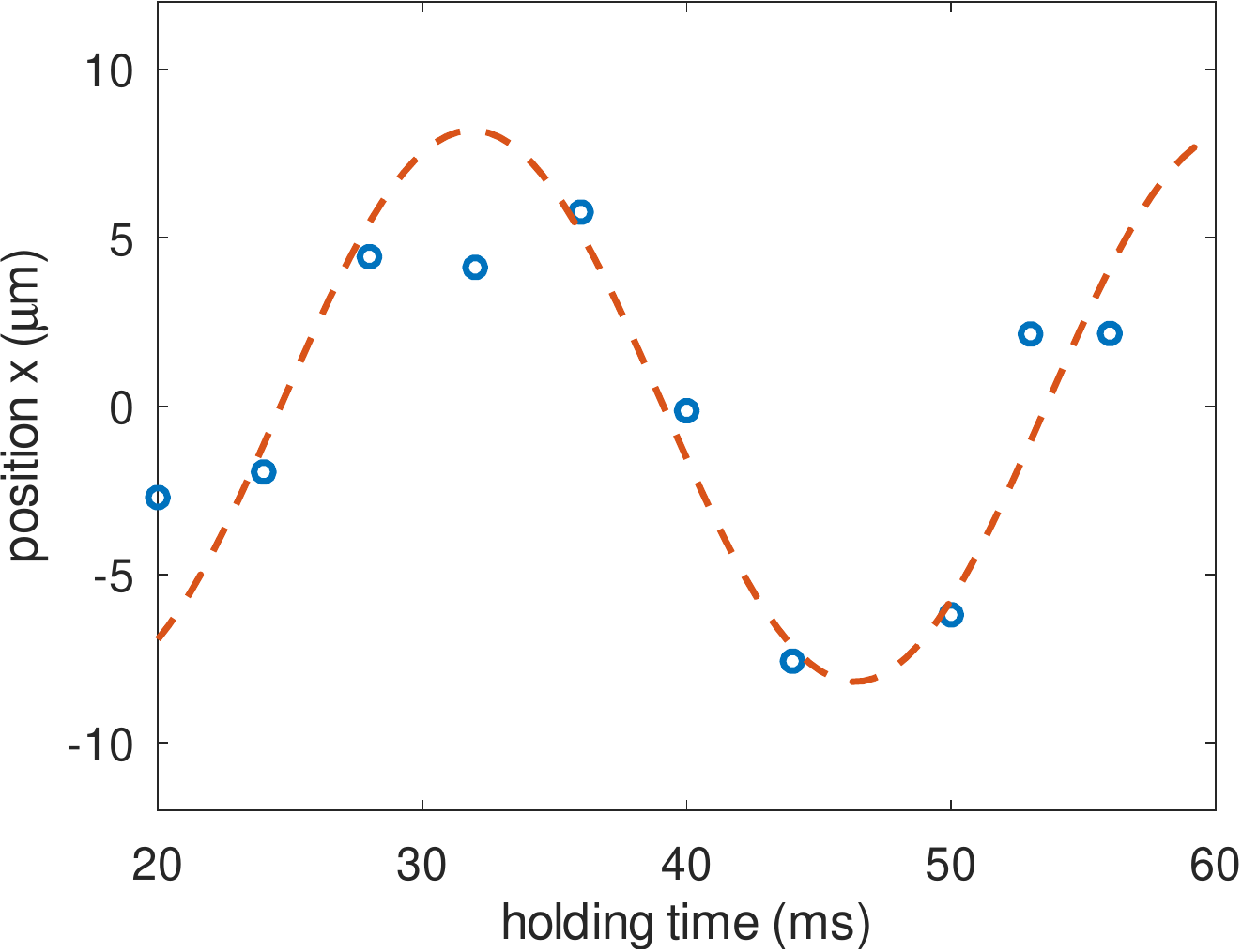}
    \caption{Residual oscillations along the $x$ axis in the shell trap, as measured after a \SI{23}{\milli\second} time-of-flight expansion. Given the oscillation frequency of \SI{34}{\hertz} in the shell trap, the oscillation amplitude of $\pm \SI{8}{\micro\metre}$ after expansion corresponds to in-trap oscillations of amplitude $\pm \SI{1.6}{\micro\metre}$.}
    \label{fig:residual_osc}
\end{figure}

The horizontal position of the laser beam becomes important when the cloud is subsequently transferred to the shell trap. Without additional optimization, the residual amplitude of the center-of-mass motion in the dressed trap is already below $\pm \SI{5}{\micro\metre}$. Optimizing the position of the dimple laser will minimize this excitation. To achieve this, we simply check the cloud properties after the transfer: atom number, temperature, and center-of-mass position. Importantly, these quantities can be obtained with a single run of the experiment, which allows for a reasonable optimization time. The final residual excitation amplitude in the direction $x$ (Fig.~\ref{fig:residual_osc}), where the trap is shallower, is below $\pm \SI{2}{\micro\metre}$, much less than the extension of the quantum gas whose Thomas--Fermi radius is \SI{16}{\micro\metre} and much better than the residual excitation observed when the atoms are loaded from a plugged quadrupole trap \cite{Merloti2013a}.

In conclusion, the preparation of a quantum gas in a shell-shaped adiabatic potential can be efficiently achieved when the atoms are initially confined in a hybrid trap where a dimple laser beam is superimposed on a quadrupole magnetic field. The cooling to degeneracy can be performed by pure rf evaporation, leading to a simple protocol. The adiabatic transfer to the shell trap is realized with simple linear ramps, leading to a condensate at rest at the bottom of the shell trap with negligible center-of-mass excitation. These results improve the state of the art in the control of sample preparation, leading to an ideal starting point for experiments of superfluid dynamics or thermodynamics in shell traps.

\section*{dedication}
This work is dedicated to the memory of Claudine Hermann (1945--2021), French physicist who was the first female professor at \'Ecole polytechnique---while mother of three sons. She was the founder and first president of the French association `Women \&\ Sciences', promoting female scientists and encouraging girls to embark on scientific careers. She was a model for a whole generation of female physicists.

\begin{acknowledgments}
LPL is UMR 7538 of CNRS and Sorbonne Paris Nord University.
This work has been supported by Région Île-de-France in the framework of DIM SIRTEQ (project Hydrolive), and by USP-COFECUB (project Uc Ph 177/19: Out of equilibrium trapped superfluids).
\end{acknowledgments}

%
\end{document}